\documentclass[12pt]{article}

\title{Generalizing the Heisenberg uncertainty relation}

\author{Eric D.\ Chisolm$^{a)}$ \\ T-1, MS B221 \\ Los Alamos National 
        Laboratory \\ Los Alamos, NM~~87545}

\begin{document}

\maketitle

\vspace*{-3.3in} \begin{flushright} LA-UR-00-1318 \end{flushright}
\vspace*{2.8in}


\begin{abstract}
The proof of the Heisenberg uncertainty relation is modified to
produce two improvements: (a) the resulting inequality is stronger
because it includes the covariance between the two observables, and
(b) the proof lifts certain restrictions on the state to which the
relation is applied, increasing its generality.  The restrictions
necessary for the standard inequality to apply are not widely known,
and they are discussed in detail.  The classical analog of the
Heisenberg relation is also derived, and the two are compared.
Finally, the modified relation is used to address the apparent paradox
that eigenfunctions of the $z$ component of angular momentum $L_z$ do
not satisfy the $\phi-L_z$ Heisenberg relation; the resolution is that
the restrictions mentioned above make the usual inequality
inapplicable to these states.  The modified relation does apply,
however, and it is shown to be consistent with explicit calculations.
\end{abstract}
{\bf I. INTRODUCTION}

The Heisenberg uncertainty relation in its general form for 
observables $A$ and $B$,
\begin{equation}
\Delta A \, \Delta B \geq \frac{1}{2} | i\langle [A, B] \rangle |,
\end{equation}
is proved in every intermediate quantum mechanics textbook (and also
in the Appendix); its best known special case, $\Delta x \, \Delta p
\geq \frac{\hbar}{2}$, comes from the canonical commutation relation
$[x, p] = i\hbar$.  A very slight modification of a standard proof 
of this inequality used by both Bohm$^1$ and Sakurai$^2$ yields two useful
improvements:
\begin{enumerate}
\item The resulting inequality is a stronger one that incorporates the 
      covariance between $A$ and $B$, a measure of their statistical 
      correlation.  As a bonus, this allows a comparison with the 
      corresponding classical inequality, in which the covariance also 
      appears.
\item This result lifts certain restrictions that must be imposed on the 
      \mbox{state} of the system for the standard Heisenberg inequality to 
      be valid.  These restrictions are not generally mentioned in textbooks, 
      but you ignore them at your peril.  For example, the $z$ component of
      angular momentum $L_z$ and the azimuthal angle $\phi$ form a
      canonical pair, so from $[\phi, L_z] = i\hbar$ one expects to
      find $\Delta \phi \, \Delta L_z \geq \frac{\hbar}{2}$.  However,
      consider the state
      \begin{equation}
      \psi(\phi) = \frac{1}{\sqrt{2\pi}} e^{im\phi}.
      \end{equation}
      This is an eigenstate of $L_z$, so $\Delta L_z = 0$, and a quick 
      calculation yields $\Delta \phi = \frac{\pi}{\sqrt{3}}$, so 
      \begin{equation}
      \Delta \phi \, \Delta L_z = 0 < \frac{\hbar}{2}.
      \end{equation}
      What went wrong?  This example has produced a flurry of commentary over 
      the years$^{3-8}$, and its resolution lies in the surprising fact that 
      eigenstates of $L_z$ do not satisfy the criteria necessary for the 
      standard Heisenberg principle to apply.  I will describe these criteria 
      in detail below, as well as why eigenstates of $L_z$ do not satisfy them,
      and once I have derived the modified inequality I will show that it is 
      consistent with this example.
\end{enumerate}
The extension to include the covariance is not new$^{9-11}$ (in fact,
it was known to Schr\"{o}dinger$^{12}$ and has been discussed before
in this journal$^{13}$), nor is the modification that removes certain
restrictions on the states$^{14,15}$.  However, the proof presented
here yields both improvements simultaneously with great ease, and the
two together allow one to discuss issues that make it clear that
quantum mechanics is not a straightforward generalization of classical
statistics, even once one has taken into account the noncommutivity of
observables.  Certain uniquely quantum mechanical concerns require
that even the definitions of statistical quantities be made with care,
as will be shown below. \\ \\
{\bf II. THE CLASSICAL UNCERTAINTY RELATION}

Since the modified inequality allows me to compare the Heisenberg relation 
with its classical counterpart, I will derive the classical relation first.
(This relation is also derived in Ref.\ 13.)

Let $a$ be a classical statistical variable with mean $\langle a \rangle$ and 
uncertainty $\Delta a$ defined by
\begin{equation}
(\Delta a)^2 = \langle ( a - \langle a \rangle)^2 \rangle = \langle a^2 
               \rangle - \langle a \rangle^2,
\end{equation}
and let $\sigma_{ab}$, the covariance between variables $a$ and $b$, be defined
by
\begin{equation}
\sigma_{ab} = \langle ( a - \langle a \rangle ) ( b - \langle b \rangle ) 
    \rangle = \langle ab \rangle - \langle a \rangle \langle b \rangle. 
\end{equation}
Notice that $(\Delta a)^2 = \sigma_{aa}$ and that $a$ and $b$ are statistically
uncorrelated if and only if $\sigma_{ab} = 0$.  I define a new variable 
$\bar{a}$ by $\bar{a} = a - \langle a \rangle$ and similarly for $b$; then 
$\langle \bar{a} \rangle = \langle \bar{b} \rangle = 0$ and 
\begin{equation}
(\Delta a)^2 = \langle \bar{a}^2 \rangle {\rm \ \ \ and \ \ \ } \sigma_{ab} = 
               \langle \bar{a} \bar{b} \rangle.
\label{defs}
\end{equation} 
Now I can prove the uncertainty relation.  Let $x$ be any statistical variable;
then $\langle x^2 \rangle \geq 0$ and $\langle x^2 \rangle = 0$ if and only if 
$x = 0$.  Then for the special case $x = \bar{a} + \lambda \bar{b}$ for any
$\lambda$ I have
\begin{equation}
\langle x^2 \rangle = \langle \bar{a}^2 \rangle + \lambda^2 \langle \bar{b}^2 
\rangle + 2\lambda \langle \bar{a} \bar{b} \rangle \geq 0
\end{equation}
with equality if and only if $\bar{a} + \lambda \bar{b} = 0$.  The central 
expression above is a quadratic in $\lambda$ which according to the inequality 
has at most one real root (if it had two then it would dip below the 
$\lambda$-axis and be negative).  The condition for the quadratic 
$Ax^2 + Bx + C$ to have at most one real root is $B^2 - 4AC \leq 0$, with 
equality in the case of exactly one root.  In this case the condition becomes
\begin{equation}
4 \langle \bar{a} \bar{b} \rangle^2 - 4 \langle \bar{a}^2 \rangle \langle 
\bar{b}^2 \rangle \leq 0,
\end{equation}
or in terms of (\ref{defs}),
\begin{eqnarray}
(\Delta a)^2 (\Delta b)^2 & \geq & (\sigma_{ab})^2 \nonumber \\
\Delta a \, \Delta b & \geq & |\sigma_{ab}|,
\label{classuncer}
\end{eqnarray}
with equality if and only if $\bar{a} + \lambda \bar{b} = 0$ for some 
$\lambda$.  This is the uncertainty principle for classical statistics. \\ \\
{\bf III. THE MODIFIED HEISENBERG RELATION}

Now I shall derive the corresponding quantum mechanical result.  Let
$A$ and $B$ be observables, and let states be denoted by $\psi, \chi$,
and so on.  The inner product of states $\psi$ and $\chi$ is denoted
$\langle \psi, \chi \rangle$, and the norm $\| \psi \|$ is defined by
$\| \psi \| = \sqrt{\langle \psi, \psi \rangle}$.  Finally, the
average of $A$ is defined by $\langle A \rangle = \langle \psi, A\psi
\rangle$.  (I deliberately avoid Dirac's $\langle \psi | A | \psi
\rangle$ because it obscures an important issue; see below.)

The quantum mechanical derivation cannot simply recapitulate the classical 
derivation with the appropriate letters capitalized for two reasons:
\begin{enumerate}
\item {\bf \mbox{\boldmath$A$} and \mbox{\boldmath $B$} might not commute.}\\
      Because of this, the order of the factors in the cross term in the 
      expansion of $\langle x^2 \rangle$ should be preserved.  The problem of 
      noncommutivity actually rears its head earlier, however, in the very 
      definition of covariance, and I must address that issue first.  The 
      classical definition of covariance is symmetric in $a$ and $b$ 
      ($\sigma_{ab} = \sigma_{ba}$) because $a$ and $b$ always commute, but if 
      I employed the same definition in the quantum case I would find   
      $\sigma_{AB} = \sigma_{BA} + \langle [A, B] \rangle$.  A covariance 
      symmetric in $A$ and $B$ is preferable, and the easiest way to achieve 
      this is to define
      \begin{eqnarray}
      \sigma_{AB} & = & \frac{1}{2}\langle (A - \langle A \rangle)(B - 
                        \langle B \rangle) + (B - \langle B \rangle)(A - 
                        \langle A \rangle)\rangle \nonumber \\ 
      & = & \frac{1}{2}\langle AB + BA \rangle - \langle A \rangle 
            \langle B \rangle.
      \end{eqnarray}
      Now $\sigma_{AB} = \sigma_{BA}$ and $\sigma_{AA}$ has the same form as 
      before, but this definition suffers from another awkward feature that 
      leads to the second point.
\item {\bf The domains of operators matter.}\\
      The domain of an operator $A$, or \mbox{$\cal{D}$}$(A)$, is the set of 
      all vectors $\psi$ in the system's Hilbert space such that $A\psi$ is 
      also a well-defined member of the Hilbert space.  (For more on operators 
      with restricted domains, see Refs.\ 16, 17, and 18.  For some of the 
      consequences for quantum mechanics, see Ref.\ 19.)  There are three main 
      reasons that a given $\psi$ might not be in \mbox{$\cal{D}$}$(A)$:
      \begin{enumerate}
      \item The operating prescription for $A$ is not defined for $\psi$.  For 
            example, consider the Hilbert space $L^2(R)$ and the momentum
            operator $p = \frac{\hbar}{i} \frac{d\,}{dx}$.  A necessary
            condition for $p\,\psi$ to exist is that $\psi$ is differentiable 
            almost everywhere (being defined almost everywhere is enough to 
            specify a member of $L^2(R)$); but to be in $L^2(R)$ a function 
            merely has to be square integrable, which does not imply 
            differentiability or even continuity.  This restriction, though 
            real, is of little practical interest, however, since it is 
            exceedingly rare in applications to encounter this problem.
      \item The operating prescription is well-defined, but the resulting 
            vector is not in the Hilbert space.  For example, again consider 
            $L^2(R)$ and the momentum operator $p$, and this time let $\psi(x) 
            = \sqrt{2|x|} e^{-|x|}$.  Now this $\psi$ is in $L^2(R)$ because it
            is square integrable (in fact, it is normalized), but its 
            derivative 
            \begin{equation} 
            \psi\,'(x) =  \frac{x}{|x|}\frac{e^{-|x|}}{\sqrt{2|x|}}(1-2|x|), 
            \end{equation}
            while well-defined everywhere except the origin, is not square 
            integrable.  Hence $\psi\,'$ is not in $L^2(R)$, so $\psi$ 
            is not in \mbox{$\cal{D}$}$(p)$.  (It is known that  
            \mbox{$\cal{D}$}$(p)$ is dense$^{20}$ in $L^2(R)$, so any $L^2$ 
            function is arbitrarily close to a function in 
            \mbox{$\cal{D}$}$(p)$, and this fact is important for quantum 
            mechanics.  Nonetheless, \mbox{$\cal{D}$}$(p)$ is not the whole 
            Hilbert space.)
      \item Sometimes \mbox{$\cal{D}$}$(A)$ is restricted to guarantee that $A$
            will be Hermitian.  For example, consider the space of $L^2$ 
            functions of the polar angle $\phi$ and the operator $L_z = 
            \frac{\hbar}{i} \frac{d\,}{d\phi}$.  For any two functions $\psi$ 
            and $\chi$, integration by parts shows that
            \begin{equation}
            \langle \chi, L_z\psi \rangle = \langle L_z\chi, \psi \rangle +
            \frac{\hbar}{i} [\chi^*(2\pi)\,\psi(2\pi) - \chi^*(0)\,\psi(0)].
            \end{equation}   
            Thus $L_z$ is Hermitian only if its domain is restricted to 
            functions $\psi$ such that $\psi(2\pi) = e^{i\alpha}\psi(0)$ for 
            some $\alpha$ (note that strict periodicity is not required).  As 
            innocent as this seems, this is the source of all of the problems 
            we encountered above with the usual form of the $\phi-L_z$ 
            uncertainty relation, as I will show below.
      \end{enumerate}
      This issue is the reason that I avoid Dirac's notation $\langle
      \chi | A | \psi \rangle$; that expression could mean either 
      $\langle \chi, A\psi \rangle$, which requires that $\psi$ is in 
      \mbox{$\cal{D}$}$(A)$ but leaves $\chi$ unrestricted, or $\langle 
      A\chi, \psi \rangle$ ($A$ is Hermitian), which reverses the restrictions 
      on $\chi$ and $\psi$.  The notation used here, on the other hand, is 
      unambiguous.  In the derivation of the uncertainty principle, I must 
      keep track of all of the domain requirements imposed on the states 
      in the proof at each step, because the final result will apply
      only to those states that satisfy {\em all} of the restrictions 
      encountered at every step.
\end{enumerate}

With these concerns in mind, I will now consider the quantum mechanical 
definitions of $\Delta A$ and $\sigma_{AB}$.  One usually defines $\Delta A$ 
by
\begin{equation}
(\Delta A)^2 = \langle \psi, (A - \langle A \rangle)^2\psi \rangle 
             = \langle A^2 \rangle - \langle A \rangle^2,
\label{olduncerdef}
\end{equation}
but notice that this expression is defined only for those states that
lie in \mbox{$\cal{D}$}$(A^2)$.  (Membership in \mbox{$\cal{D}$}$(A)$
is a prerequisite for membership in \mbox{$\cal{D}$}$(A^2)$.)  Now I
would certainly like $\Delta A$ to be defined for every state for
which $\langle A \rangle$ is defined, so I'd like $\Delta A$ to exist
for every state in \mbox{$\cal{D}$}$(A)$.  The easiest way to do this
is to note that by the Hermiticity of $A$, for all states for which
the above definition is valid it is equivalent to
\begin{equation}
(\Delta A)^2 = \langle  (A - \langle A \rangle)\psi, (A - \langle A \rangle)
               \psi \rangle = \| (A - \langle A \rangle)\psi \|^2,
\label{uncerdef}
\end{equation}
and {\em this} expression is defined for every state in \mbox{$\cal{D}$}$(A)$.
Hence I take Eq.\ (\ref{uncerdef}), not Eq.\ (\ref{olduncerdef}), to be my 
definition for $\Delta A$.  Remember that it is equivalent to the old 
definition whenever the old definition is valid, but the old definition is not 
valid in every case where I would like it to be.

Now on to $\sigma_{AB}$.  The definition suggested above,
\begin{eqnarray}
\sigma_{AB} & = & \frac{1}{2}\langle \psi, [\,(A - \langle A \rangle)(B - 
                  \langle B \rangle) + (B - \langle B \rangle)(A - 
                  \langle A \rangle)\,]\psi \rangle \nonumber \\ 
& = & \frac{1}{2}\langle \psi, (AB + BA)\psi \rangle - \langle A \rangle 
                  \langle B \rangle,
\label{oldcovdef}
\end{eqnarray}
requires that both $AB\psi$ and $BA\psi$ exist, or that $\psi$ is in both  
\mbox{$\cal{D}$}$(AB)$ and \mbox{$\cal{D}$}$(BA)$.  However, I would prefer a 
definition of $\sigma_{AB}$ that made only the weaker requirement that $\psi$ 
is in both \mbox{$\cal{D}$}$(A)$ and \mbox{$\cal{D}$}$(B)$, not least because 
I want to relate $\sigma_{AB}$ to $\Delta A$ and $\Delta B$, and the weaker 
requirement is all that is needed to guarantee their existence.  Fortunately, 
this is easy; the Hermiticity of $A$ and $B$ allows me to rewrite the above as
\begin{eqnarray}
\sigma_{AB} & = & \frac{1}{2}\langle (A - \langle A \rangle)\psi, (B - 
                  \langle B \rangle)\psi \rangle + \frac{1}{2}\langle (B - 
                  \langle B \rangle)\psi, (A - \langle A \rangle)\psi \rangle
                  \nonumber \\
& = & {\rm Re}\langle (A - \langle A \rangle)\psi, (B - \langle B \rangle)\psi 
              \rangle \nonumber \\
& = & {\rm Re}\langle A\psi, B\psi \rangle - \langle A \rangle 
              \langle B \rangle,
\label{covdef}
\end{eqnarray}
and this definition is valid on the larger set of states that belong
to both \mbox{$\cal{D}$}$(A)$ and \mbox{$\cal{D}$}$(B)$, exactly as
desired.  Hence I take Eq.\ (\ref{covdef}), not Eq.\ (\ref{oldcovdef}), 
as the definition of covariance.  Again, the two expressions are equivalent 
whenever both are defined, but the first does not exist in every case 
where I would like it to be, whereas the second does.  Finally, in analogy 
with the classical case I define $\bar{A} = A - \langle A \rangle$, in 
terms of which
\begin{equation}
\Delta A = \| \bar{A}\psi \| {\rm \ \ \ and \ \ \ } 
\sigma_{AB} = {\rm Re}\langle \bar{A}\psi, \bar{B}\psi \rangle.
\label{newuncercov}
\end{equation}
Note that $(\Delta A)^2 = \sigma_{AA}$, just as in the classical case.

Now for the uncertainty relation.  The Cauchy-Schwarz inequality says that for 
any states $\psi$ and $\chi$,
\begin{equation}
| \langle \chi, \psi \rangle | \leq \| \chi \| \, \| \psi \|.
\end{equation}
Then, using Eq. (\ref{newuncercov}),
\begin{eqnarray}
\Delta A \, \Delta B & = & \|\bar{A}\psi\| \, \|\bar{B}\psi\| \nonumber \\
            & \geq & | \langle \bar{A}\psi, \bar{B}\psi \rangle | \nonumber \\
            & = & \sqrt{({\rm Re}\langle \bar{A}\psi, \bar{B}\psi \rangle)^2 + 
                        ({\rm Im}\langle \bar{A}\psi, \bar{B}\psi \rangle)^2}
                        \nonumber \\
            & = & \sqrt{\sigma_{AB}^2 + ({\rm Im}\langle 
                              \bar{A}\psi, \bar{B}\psi \rangle)^2}.
\end{eqnarray}
A little algebra shows that Im$\langle \bar{A}\psi, \bar{B}\psi 
\rangle =$ Im$\langle A\psi, B\psi \rangle$, so the final result is
\begin{equation}
\Delta A \, \Delta B \geq \sqrt{\sigma_{AB}^2 + ({\rm Im}\langle A\psi, B\psi 
\rangle)^2}.
\label{Heisenberg}
\end{equation}
This is the modified Heisenberg uncertainty relation.\\ \\
{\bf IV. COMMENTS}

First, note that all of the steps leading to Eq.\ (\ref{Heisenberg}) are valid 
as long as $\psi$ lies in both \mbox{$\cal{D}$}$(A)$ and \mbox{$\cal{D}$}$(B)$,
and consequently so is the final result.  Therefore, unlike the usual form of 
the Heisenberg relation, this inequality is guaranteed to hold in all 
circumstances in which the quantities involved (the uncertainties and 
covariances) are well-defined; there are no more unpleasant surprises waiting 
to be discovered.

Next, I shall recover the uncertainty relation with which we are familiar.  
If $\psi$ lies in both \mbox{$\cal{D}$}$(AB)$ and \mbox{$\cal{D}$}$(BA)$, then
the following manipulations are allowed:
\begin{eqnarray}
{\rm Im}\langle A\psi, B\psi \rangle & = & -\frac{i}{2}\langle A\psi, B\psi 
      \rangle + \frac{i}{2}\langle B\psi, A\psi \rangle \nonumber \\
& = & -\frac{i}{2}\langle \psi, AB\psi \rangle + \frac{i}{2}\langle \psi, 
      BA\psi \rangle \nonumber \\
& = & -\frac{i}{2}\langle \psi, (AB-BA)\psi \rangle \nonumber \\
& = & -\frac{i}{2}\langle [A, B]\rangle.
\end{eqnarray}
Thus when this additional condition is satisfied,
\begin{equation}
\Delta A \, \Delta B \geq \sqrt{\sigma_{AB}^2 + 
                          \frac{1}{4}(i\langle[A, B]\rangle)^2},
\label{modHeisenberg}
\end{equation}
which implies the standard Heisenberg inequality.

Comparing Eq.\ (\ref{classuncer}) with either (\ref{Heisenberg}) or 
(\ref{modHeisenberg}), we see that the sole difference introduced by quantum 
mechanics is the term Im$\langle A\psi, B\psi \rangle$, which on a fairly 
large class of states is essentially half the expectation value of $i$ times 
the commutator $[A, B]$.  This is the irreducible indeterminacy present even in
states where the two observables are entirely independent statistically.

Now I can reconsider the example of the $\phi-L_z$ uncertainty relation 
discussed at the beginning.  For the commutator form of the inequality to 
apply, $\psi$ must lie in the domains of both $\phi\,L_z$ and $L_z\,\phi$, and 
$\psi = (2\pi)^{-1/2}\exp(im\phi)$ does not satisfy the latter criterion.
If it did, then that would mean that $\phi\,\psi$ would be in the domain of 
$L_z$, but as I noted earlier every state in the domain of $L_z$ must satisfy 
$\psi(2\pi) = e^{i\alpha} \psi(0)$, and 
\begin{equation}
\phi \psi(\phi) = \frac{\phi}{\sqrt{2\pi}} e^{im\phi}
\end{equation}
vanishes at $\phi = 0$ and is nonvanishing at $\phi = 2\pi$.  Hence $(L_z\,
\phi)\psi$ does not exist, and the commutator inequality does not apply.
However, Eq.\ (\ref{Heisenberg}) does apply, and to find it for this special 
case I calculate
\begin{eqnarray}
{\rm Im}\langle \phi\,\psi, L_z\,\psi \rangle & = & -\frac{i}{2}\langle \phi\,
      \psi, L_z\,\psi \rangle + \frac{i}{2}\langle L_z\,\psi, \phi\,\psi 
      \rangle \nonumber \\
& = & -\frac{\hbar}{2} \int_{0}^{2\pi} \phi\,\psi^*\,\frac{d\psi}{d\phi}\,d\phi
      -\frac{\hbar}{2} \int_{0}^{2\pi} \frac{d\psi^*}{d\phi}\,\phi\psi\,d\phi
      \nonumber \\
& = & -\frac{\hbar}{2} \int_{0}^{2\pi} \phi \left(\psi^*\,\frac{d\psi}{d\phi} +
                       \psi\,\frac{d\psi^*}{d\phi} \right)\,d\phi \nonumber \\
& = & -\frac{\hbar}{2} \int_{0}^{2\pi} \phi\,\frac{d\,}{d\phi}(\psi^*\psi) 
      \, d\phi \nonumber \\
& = & -\frac{\hbar}{2}\,[\phi\,\psi^*\psi]^{2\pi}_{0} + \frac{\hbar}{2} 
      \int_{0}^{2\pi} \psi^*\psi \, d\phi \nonumber \\
& = & \frac{\hbar}{2}\left(1 - 2\pi|\psi(2\pi)|^2\right). 
\end{eqnarray}
Thus
\begin{equation}
\Delta \phi \, \Delta L_z \geq \sqrt{ \sigma_{\phi L_z}^2 + \frac{\hbar^2}{4}
                               \left(1 - 2\pi|\psi(2\pi)|^2\right)^2}.
\label{phiLrelation}
\end{equation}
For the particular $\psi$ in question, $\sigma_{\phi L_z} = 0$ (again because 
$\psi$ is an eigenstate of $L_z$) and $|\psi(2\pi)|^2 = (2\pi)^{-1}$, so
\begin{equation} \Delta \phi \, \Delta L_z \geq 0, \end{equation}
which is consistent with what we found at the beginning.

Incidentally, if one carried out an analogous derivation with $x$ and $p$ in 
place of $\phi$ and $L_z$, one would find
\begin{equation}
{\rm Im}\langle x\psi, p\psi \rangle = \frac{\hbar}{2}\left(1 - 
                                       [x\psi^*\psi]^{\infty}_{-\infty}\right),
\end{equation}
so the usual Heisenberg inequality for $x$ and $p$ is valid as long as
$\psi$ \mbox{falls} off faster than $|x|^{-1/2}$ as $|x| \rightarrow \infty$.
Since $\psi$ is differentiable almost everywhere it must fall off smoothly, 
in which case square integrability imposes the above requirement 
automatically.  Hence the standard form of the Heisenberg inequality is always 
valid for $x$ and $p$.  It is precisely the fact that the coordinate $\phi$ is 
bounded while $x$ is unbounded that allows the sorts of problems considered in 
this paper to crop up often in one case and not at all in the other.  

One final note is in order concerning the $\phi-L_z$ inequality.  In its 
current form, Eq.\ (\ref{phiLrelation}), the inequality is not invariant under 
rotations, as one would prefer, since the direction corresponding to $\phi = 0$
has no physical significance.  (The fact that one must choose a $\phi = 0$ 
direction just to define $\phi$ is the source of the problem.)  Hence the 
$\phi-L_z$ inequality has still not been brought to a quite satisfactory form; 
to finish the job, one must develop rotation-invariant definitions of 
uncertainty and repeat the proof, which has been done in Ref.\ 4. \\ \\
{\bf APPENDIX:  ANOTHER STANDARD PROOF OF THE HEISENBERG RELATION}

This proof of the uncertainty relation is found, for example, in Ref.\ 21.
Let $A$ and $B$ be observables, let $\psi$ be a state in both 
\mbox{$\cal{D}$}$(AB)$ and \mbox{$\cal{D}$}$(BA)$ (and thus in 
\mbox{$\cal{D}$}$(A)$ and  \mbox{$\cal{D}$}$(B)$), and let $\bar{A}$ and 
$\bar{B}$ be defined as earlier.  Then for any real $\lambda$
\begin{eqnarray}
\|(\bar{A} + i\lambda\bar{B})\psi\|^2 & \geq & 0 \nonumber \\
\langle \psi, (\bar{A} - i\lambda\bar{B})(\bar{A} + i\lambda\bar{B})\psi 
                              \rangle & \geq & 0 \nonumber \\
\langle \psi, (\bar{A}^2 + \lambda^2\bar{B}^2 + i\lambda[\bar{A}\bar{B} - 
         \bar{B}\bar{A}])\psi \rangle & \geq & 0 \nonumber \\
(\Delta A)^2 + \lambda^2(\Delta B)^2 + i\lambda \langle [A, B] \rangle & 
                                                    \geq & 0,
\end{eqnarray}
where the last line used the standard quantum mechanical definition of 
uncertainty and the fact that $[\bar{A}, \bar{B}] = [A, B]$.  The commutator of
two observables is anti-Hermitian, so the quantity $i\langle [A, B] \rangle$ is
real.  Again we have a quadratic in $\lambda$ with at most one real root, so 
the same condition as mentioned in the text yields
\begin{equation}
(i\langle [A, B] \rangle)^2 - 4(\Delta A)^2(\Delta B)^2 \leq 0,
\end{equation}
or  
\begin{equation}
\Delta A\,\Delta B \geq \frac{1}{2}|i\langle [A, B] \rangle|.
\end{equation}
This is the standard Heisenberg uncertainty relation.  This result can be 
strengthened by replacing $i\lambda$ with $\lambda e^{i\theta}$, treating 
$\lambda$ as before, and taking the maximum over all $\theta$; the result is 
Eq.\ (\ref{modHeisenberg}).  If one modifies this derivation to take into 
account the new definitions of $\Delta A$ and $\sigma_{AB}$, Eq.\ 
(\ref{newuncercov}), one recovers the main result of this paper, Eq.\ 
(\ref{Heisenberg}).  The derivation in Sec.\ III is much shorter, however.
\vspace{.5in}
\begin{description}
\item[a)] Electronic mail:  echisolm@lanl.gov
\item[1.] D.\ Bohm, {\it Quantum Theory} (Prentice-Hall, Englewood Cliffs, NJ, 
          1951), pp.\ 205-207.
\item[2.] J.\ J.\ Sakurai, {\it Modern Quantum Mechanics} rev.\ ed.\ 
          (Addison-Wesley, New York, 1994), pp.\ 34-36.
\item[3.] D.\ Judge, ``On the uncertainty relation for angle variables,'' 
          Nuovo Cimento {\bf 31} (1964), pp.\ 332-340. 
\item[4.] K.\ Kraus, ``Remark on the uncertainty between angle and angular 
          momentum,'' Z.\ Phys.\ {\bf 188} (1965), pp.\ 374-377.
\item[5.] P.\ Carruthers and M.\ M.\ Nieto, ``Phase and angle variables in 
          quantum mechanics,'' Rev.\ Mod.\ Phys.\ {\bf 40} (1968), pp.\ 
          411-440.
\item[6.] J.\ M.\ Levy-Leblond, ``Who is afraid of nonhermitian operators?  
          A quantum description of angle and phase,'' Ann.\ Phys.\ NY {\bf 101}
          (1976), 319-341.
\item[7.] F.\ Gesztesy and L.\ Pittner, ``Uncertainty relations and quadratic 
          forms.'' J.\ Phys.\ A {\bf 11} (1978), pp.\ 1765-1770.
\item[8.] A.\ Galindo and P.\ Pascual, {\it Quantum Mechanics} (Springer-Verlag, 
          New York, 1990), vol.\ 1, pp.\ 201-206.
\item[9.] C.\ W.\ Gardiner, {\it Quantum Noise} (Springer-Verlag, New York, 
          1991), pp.\ 1-2.
\item[10.]V.\ V.\ Dodonov, E.\ V.\ Kurmyshev, and V.\ I.\ Man'ko, ``Generalized
          Uncertainty Relation and Correlated Coherent States,'' Phys.\ 
          Lett.\ A {\bf 79} (1980), pp.\ 150-152.  This paper also shows that 
          the uncertainty relation with covariance is valid for mixed states as
          well.  
\item[11.]V.\ V.\ Dodonov and V.\ I.\ Man'ko, ``Generalization of the 
          Uncertainity Relations in Quantum Mechanics,'' in {\it Invariants and
          the Evolution of Nonstationary Quantum Systems}, Vol.\ 183 of the 
          Proceedings of the Lebedev Physics Institute, ed.\ M.\ A.\ Markov 
          (Nova Science, Commack, NY, 1989), pp.\ 3-101.  This article also 
          discusses the $\phi-L_z$ uncertainty relation as well as 
          higher-order uncertainty relations, relations among an arbitrary 
          number of observables, and entropy-based uncertainty relations, and 
          it has a substantial list of references.  
\item[12.]E.\ Schr\"{o}dinger, ``Zum Heisenbergschen unsch\"{a}rfeprinzip,''
          Sitzungsber.\ K.\ Preuss.\ Akad.\ Wiss.\ (1930), pp. 296-303.
\item[13.]J.\ Peslak, Jr., ``Comparison of classical and quantum mechanical 
          uncertainties,'' Am.\ J.\ Phys.\ {\bf 47} (1979), pp.\ 39-45.  
\item[14.] F.\ Gieres, ``Dirac's formalism and mathematical surprises in 
          quantum mechanics,'' quant-ph/9907069, pp.\ 21-22, 26-28. 
\item[15.]See Refs.\ 4 and 7.
\item[16.]N.\ I.\ Akhiezer and I.\ M.\ Glazman, {\it Theory of Linear Operators
          in Hilbert Space} (Ungar, New York, 1961 and 1963), 2 vol.
\item[17.]N.\ Dunford and J.\ T.\ Schwartz, {\it Linear Operators} 
          (Interscience, New York, 1958, 1963, and 1971), 3 vol.
\item[18.]F.\ Riesz and B.\ Sz.-Nagy, {\it Functional Analysis} (Ungar, New 
          York, 1955).
\item[19.]J.\ M.\ Jauch, {\it Foundations of Quantum Mechanics} 
          (Addison-Wesley, New York, 1968). 
\item[20.]See Ref.\ 19, p.\ 43.
\item[21.]C.\ Cohen-Tannoudji, B.\ Diu, and F.\ Lalo\"{e}, {\it Quantum 
          Mechanics} (John Wiley and Sons, New York, 1977), Vol.\ 1, pp. 
          286-287.
\end{description}

\end{document}